\documentclass[12pt,preprint]{aastex}
\usepackage{amstext}
\usepackage{apjfonts}
\usepackage{amsmath}
\usepackage{psfig}

\shorttitle{X-ray emission iron line in GRB X-ray afterglows}
\shortauthors{W.H.Gao \& D.M.Wei}

\begin{document}

\title{A NEW MODEL FOR IRON EMISSION LINES AND RE-BURST IN GRB X-RAY AFTERGLOWS}

\author{W.H.Gao and D.M.Wei}
\affil{Purple Mountain Observatory,Chinese Academy of
Science,Nanjing,210008,China  \\
 National Astronomical Observatories,Chinese Academy of
Science,China }

\begin{abstract}
Recently iron emission features have been observed in several
X-ray afterglows of GRBs. It is found that the energy obtained
from the illuminating continuum which produces the emission lines
is much higher than that of the main burst.The observation of
SN-GRB association indicates a fallback disk should be formed
after the supernovae explosion. The disk is optically thick and
advection-dominated and dense. We suggest that the delayed
injection energy after the initial main burst, much higher than
energy of the main burst, causes the re-burst appearance in GRB
afterglow and illuminates the region of the disk surface with
$\tau\approx1$($\tau$ is the optical depth for the Thomson scatter
) and produces the iron emission line whose luminosity can be up
to $10^{45}$ erg$s^{-1}$. The duration of the iron line emission
can be $10^{4}-10^{5}$ s. This model can explain the appearance of
re-burst and emission lines in GRB afterglow and disappearance of
the iron emission lines, and also can naturally solve the problem
of higher energy of the illuminating continuum than that of the
main burst. This scenario is different from the models put forward
to explain the emission lines before, that can be tested by SWIFT
satellite.

\end{abstract}


\section{Introduction}
Recently evidence is mounting that 'long duration' GRBs are
associated with a rare type of supernovae event such as a 'failed
supernova','Hypernova'or 'Collapsar'( Woosley 1993; Paczy\'{n}ski
1998; MacFadyen \& Woosley 1999; MacFadyen,Woosley and Herger
2001; Proga et al. 2003). In this model, the time between the
supernovae explosion and the GRB is very short, nearly
simultaneous. Another model called 'Supranova model' was proposed
by Vietri $\&$ Stella (1998), in which the supernova(SN) explosion
initially results in the formation of a comparatively massive,
magnetized neutron star endowed with rapid rotation. This
supramassive neutron star is envisioned to gradually lose
rotational support through a pulsar-type wind until it eventually
becomes unstable to gravitational collapse,leading to the
formation of a BH and the triggering of a GRB. In 'Supranova
model', the time between the supernovae explosion and the GRB can
range from several weeks to several years.

After the supernovae explosion, the fallback material will form a
disk around the newly central black hole in either 'collapsar
model' or 'supranova model', and this disk is advection-dominant
and hot and dense even after the GRBs(e.g. Chevalier 1989;
Mineshige et al. 1997).

X-ray emission lines observed in X-ray afterglow of GRBs provide
important clues for identifying the nature of the progenitors of
long (t$\geq$2 s) GRBs. The first marginal detection of an
emission line was in the X-ray afterglow of GRB 970508 with the
BeppoSAX NFI (Piro et al. 1999). Later emission lines were also
detected in the X-ray afterglows of GRB 970828 (Yoshida et al.
2001) with ASCA; GRB 991216 (Piro et al. 2000) and GRB020813
(Butler et al. 2003) with Chandra; GRB 011211 (Reeves et al.
2002), GRB 001025A (Watson et al. 2002) and GRB 030227 (Watson et
al. 2003) with XMM-Newton; GRB 000214 (Antonelli et al. 2000) with
BeppoSAX. The detailed properties of the X-ray emission features
can be found in several papers ( Lazzati 2002; B\"{o}ttcher 2003;
Gao \& Wei 2004 ). The energy of the emission line found in the
X-ray afterglows of GRB 970508, GRB 970828, GRB 991216 and GRB
000214 is roughly consistent with Fe $K_{\alpha}$ at the redshift
of the hosts.It is adduced as evidence that the environment of the
burst is heavily enriched in iron as the result of a recent
supernova explosions ( e.g. Lazzati et al. 1999; Ghisellini et al.
1999). We would have observed Fe line if the time delay is more
than several months such as models of Supranova (Vietri \& stella
1998).

The standard model for gamma-ray burst afterglows assumes that
relativistic material is decelerating on account of interaction
with the surrounding medium,with a nearly impulsive injection
energy. But perhaps the ejecta consists of many concentric shells
moving at different speeds, slow moving material carries most of
the system's energy. The delayed injection energy can be more than
that of the main burst. It is proposed to be refreshed shock
scenario. In the refreshed shock scenario, assuming the source
ejects a range of Lorentz factors with the mass
$M(>\gamma)\propto \gamma^{-s}$ (Rees $\&$
M$\acute{e}$sz$\acute{a}$ros 1998; Sari \&
M$\acute{e}$sz$\acute{a}$ros 2000).

The re-burst phenomena have been found in the afterglow of
GRB970508(Piro et al. 1998) and GRB970828(Yoshida et al. 1999). It
has been considered that energy of delayed injection with more
than that of the initial burst caused the re-burst
appearance(Panaitescu et al. 1998; Kumar $\&$ Piran 2000; Sari
$\&$ M$\acute{e}$sz$\acute{a}$ros 2000).

The energy contained in the illuminating continuum that is
responsible for the line production is much higher than that of
the collimated GRBs (Lazzati 2002; Ghisellini et al. 2002; Gao \&
Wei 2004). In this letter we consider the illuminating continuum
which produces the emission line comes from the delayed injection
energy , which illuminates the fallback disk and produces the
emission lines.

\section{Delayed energy injection and Fe emission line feature in X-ray afterglow}

The re-burst of emission during the afterglow has been reported in
two GRBs,GRB970508 (Piro et al. 1998) and GRB970828 (Yoshida et
al. 1999). At the same time possible evidence for the existence of
Fe $K\alpha$ line has also been found in these two GRBs
(GRB970508, Piro et al.1999; GRB970828, Yoshida et al. 1999).
Delayed energy injection (or refreshed shock model) can well
explained the resurgence emission in these two GRBs (Panaitescu et
al. 1998; Kumar \& Piran 2000). The delayed injection energy more
than the initial fireball energy can produce the observed re-burst
in the afterglow about 0.5 day after $\gamma$-ray burst. Here we
model that the delayed injection energy illuminates the fallback
disk around the central black hole and photonionizes the layer of
the disk with $\tau$=1, and iron line emission can be produced by
the recombination process.

Considering an engine that emits both an initial impulsive energy
input as well as a continuous luminosity, the latter varying as a
power law in the emission time. The differential energy
conservation relation in the observer's frame can be expressed as
(Cohen \& Piran 1999; Zhang \& M$\acute{e}$sz$\acute{a}$ros 2001)
\begin{equation}
dE/dt_{\oplus}=L_{0}(t_{\oplus}/t_{0})^{q}-k(E/t_{\oplus}) ,
t_{\oplus}>t_{0} .
\end{equation}

The first term on the right side is
L=$L_{0}(t_{\oplus}/t_{0})^{q}$ which is the intrinsic luminosity
of refresh shock, $t_{0}$ is a characteristic timescale for the
formation of a self-similar solution, E and $t_{\oplus}$ denote
the energy and time measured in the observer frame, q and k are
dimensionless constants.

In the refreshed shock scenario, following Rees $\&$
M$\acute{e}$sz$\acute{a}$ros (1998), Kumar $\&$ Piran(2000), Sari
$\&$ M$\acute{e}$sz$\acute{a}$ros (2000), one can obtain the
relationship between the temporal index $\alpha$ and the spectral
index $\beta$, where $F_{\nu}\propto t_{\oplus}^{\alpha}
\nu^{\beta}$. For the X-ray afterglow, considering the forward
shock in the slow-cooling regime(Sari, Piran, \& Narayan 1998),
$\alpha$=(1-q/2)$\beta$+1+q (Zhang \&
M$\acute{e}$sz$\acute{a}$ros,2001).  Since no spectral information
was available for GRB970508 in the first several days of the
afterglow, in the calculation of Panaitescu et al. (1998), s=1.5
is needed. For the forward shock in the slow-cooling regime in the
refreshed model, $\alpha=-\frac{6-6s-24\beta}{2(7+s)}$(Sari \&
M$\acute{e}$sz$\acute{a}$ros 2000, for the uniform medium
environment). So $\beta$=(17$\alpha$-3)/24. In the X-ray afterglow
observations of GRB970508, the temporal index $\alpha$ changes
from -1.1 (before the re-burst) to +1.7 (at the beginning of the
re-burst) to -0.4 (during the re-bust) and then to -2.2 (after the
re-burst) (Piro et al. 1999).

From above in the refreshed shock scenario one can obtain the
delayed energy injection as L$\propto t_{\oplus}^{q}$,
q$\sim$-0.8, which ends after about $10^{5}$s. We assume the
delayed energy is isotropic, that is reasonable from the
observations. On the observational side, Pedersen et al.(1998)
have found the optical light curve of GRB970508 afterglow can be
explained in terms of an isotropic outflow. The radio observation
of GRB991216 afterglow also showed there was an isotropically
energetic fireball($10^{54}$erg)(Frail et al. 2000). From the
numerical work of Panaitescu et al.(1998), the injection energy
injects into the GRB outflow up to be $E_{inj}=3E_{0}$ into the
external medium, whose solid angle is $\Omega_{\gamma}$,
$\Omega_{\gamma}$ is the solid angle of the GRB collimated jet. We
name this energy as Injection Energy, that is only a fraction of
the whole delayed energy. The whole delayed energy is about
$E_{del}=\frac{4\pi}{\Omega_{\gamma}}E_{inj}$. Luminosity varies
as about $t_{\oplus}^{-0.8}$. At the beginning of the re-burst,
$t_{i}\sim 6\times 10^{4}s$, the Injection Energy has been as much
as that of the initial burst $E_{0}\sim 4 \times
10^{50}$ergs(Bloom et al. 2003). After that time, the residual
Injection Energy 2$E_{0}$ has been exhausted in about $10^{5}$
s(Piro et al. 1999). From
\begin{equation}
E=\int^{t_{e}}_{t_{i}}L_{i}(\frac{t}{t_{i}})^{-0.8}dt
\end{equation}

$t_{i}$ means the time at which the re-burst appears, $t_{e}$
means the time when the re-burst ends, $L_{i}$ is the luminosity
at the time of re-burst emergence, E means the energy injecting
into the external medium. After the emergence of the re-burst, the
residual Injection Energy is about 2$E_{0}$, and time duration is
about $10^{5}$s, we can obtain that $L_{i}\sim10^{46}ergs^{-1}$.
From $L\propto t^{-0.8}$ and about energy of  $E_{0}$ has been
exhausted before the re-burst, we can get $L\sim10^{47}erg
s^{-1}$, at $t\sim3\times10^{3}$s.

An advection-dominant disk can exist around a stellar black hole
even after the $\gamma$-ray bursts(Kohri \& Mineshige 2002;Janiuk
et al 2004)(\textbf{Fig.1}). For the 'collapsar model', SN
explodes at almost same time with the GRBs, or about a minute to
few hours prior to the $\gamma$-ray bursts. Whereas for 'supranova
model', the time delay between the SN explosion and GRB is perhaps
several months or even longer. In this case almost all the
fallback nickel has been decayed to iron.

The evolution of fallback disk around the black hole has been
considered by, eg., Meyer $\&$ Meyer-Hofmeister(1989), Cannizzo et
al.(1990), Mineshige et al.(1993), and Mineshige et al.(1997).
From the work of Mineshige et al. (1997),we can draw the accretion
rate of the disk result from the fallback material. Since the
radioactivity time scale is about 85 days for nickel decaying to
iron, here we adopt 'Supranova model'(Vietri \& Stella 1998), and
assume $\gamma$-ray bursts take place about 100 days after the SN
explosion, .
\begin{equation}
\dot{M}\sim 10^{25}g
s^{-1}(\frac{M_{BH}}{3M_{\odot}})(\frac{\Delta M}{0.1
M_{\odot}})(\frac{\alpha}{0.01})(\frac{t}{100 day})^{-1.35}
\end{equation}
where $\Delta$M is the amount of fallback material of the disk and
$\alpha$ is the viscosity parameter.We adopt that the mass of the
black hole to be 3 $M_{\odot}$. $\Delta$M has a large range for
the different mechanism of the SN explosion and the evolution of
the disk (eg. Woosley 1993; Chevalier 1989). Here we adopt
$\Delta$M to be 0.1 $M_{\odot}$.

While about one hundred days after the SN explosion, the accretion
rate would be decreasing to about $10^{-8}{M_{\odot}}$$s^{-1}$. At
this time the radiation pressure is dominant in the disk (eg.
Mineshige et al.1997). The relations of T-$\Sigma$ and
$\dot{M}$-$\Sigma$ are as following (Kohri \& Mineshige 2002):
\begin{equation}
T=4.87\times10^{1}\alpha^{0}\Sigma^{\frac{1}{4}}r^{-\frac{1}{2}}M_{BH}^{\frac{1}{4}}
\end{equation}

\begin{equation}
\Sigma=3.3\times10^{3}\dot{M}\alpha^{-1} r^{-\frac{1}{2}}
M_{BH}^{-\frac{1}{2}}
\end{equation}

The total mass of the disk is
\begin{equation}
\Delta M=\int^{R_{o}}_{R_{in}} 2 \pi R \Sigma dR
\end{equation}

$R_{in}$ and $R_{o}$ represent the innermost radius and the
outermost radius respectively. From eq.(3), eq.(5) and eq.(6),
adopting $\Delta$M=0.1$M_{\odot}$, we can obtain the radius of the
outermost disk, $R_{o}\sim 2\times10^{12}$cm.

So we can get the results about surface density $\Sigma$ and
temperature T at the outermost radius:
\begin{equation}
\Sigma=4.5\times10^{7}gcm^{-2}(\frac{\dot{M}}{10^{25}g
s^{-1}})(\frac{\alpha}{0.01})^{-1}
\end{equation}
$$\times(\frac{r}{10^{6}
r_{s}})^{-\frac{1}{2}}(\frac{M_{BH}}{3M\odot})^{-\frac{1}{2}}$$

and

\begin{equation}
T=2\times10^{6}K(\frac{r}{10^{6}r_{s}})^{-\frac{1}{2}}(\frac{M_{BH}}{3M\odot})^{\frac{1}{4}}
\end{equation}
Then the average density of the disk at this time at the outermost
radius r=$10^{6}r_{s}$ is:

\begin{equation}
\rho=\frac{\Sigma}{2H}=7.2 \times 10^{-5}
gcm^{-3}(\frac{\Sigma}{4.5\times 10^{7}g cm^{-2}})^{2}
\end{equation}
$$\times(\frac{T}{2\times10^{6}K})^{-4}(\frac{r}{10^{6}r_{s}})^{-3} (\frac{M}{3M_{\odot}})$$

At the outermost radius which is about
r=$10^{6}$$r_{s}$=8.85$\times 10^{11}$cm, temperature T is about
2$\times10^{6}$K and the number density n is about
4$\times10^{19}cm^{-3}$ in the disk.

The ionization parameter is
$\xi=\frac{L_{ill}}{nR^{2}}=2.5\times10^{3}(\frac{L_{ill}}{10^{47}ergs^{-1}})
(\frac{n}{4\times10^{19}})^{-1}(\frac{R}{10^{12}cm})^{-2}$. At
this ionization parameter, iron emission is very efficient(Lazzati
et al. 2002). The recombination time for hydrogenic iron in the
outer disk photoionized by the nonthermal delayed energy
is(Lazzati et al. 1999)
\begin{equation}
t_{rec}=1.5\times10^{-8}T_{8}^{\frac{1}{2}}n_{17}^{-1}s
\end{equation}
The temperature parametrization used here is consisitent with the
range expected from photoionization equilibrium (Lazzati et al.
1999).

The optical depth at outer radius r=$10^{12}$cm is optically
thick: $\tau$$_{T}$=2nH$\sigma_{T}$$\sim10^{7}$ . So the number of
Fe nuclei in the layer of the disk with $\tau=1$ is
N$_{Fe}$$\sim$$\chi$$_{Fe}$M/($\tau$$_{T}$56m$_{p}$)$\sim$10$^{47}$$\chi$$_{Fe}$.
($\chi$$_{Fe}$ is the iron mass fraction of the disk ). The Fe
line luminosity is [N$_{Fe}$(8 KeV)/t$_{rec}$](1+z)$^{-1}$, or

\begin{equation}
L_{Fe}\sim\chi_{Fe}10^{47}(1+z)^{-1} ergs^{-1}
\end{equation}

For SN1987A, $\chi$$_{Fe}$ can be about 2$\%$ when all the nickel
have decayed to iron(Chevalier 1989). So the luminosity of iron
line can be obtained:
$L_{Fe}$$\sim$2$\times$$10^{45}$(1+z)$^{-1}$erg$s^{-1}$.

After the emergence of the re-burst, luminosity decays from about
$10^{46}erg s^{-1}$ at the rate of $t^{-0.8}$. So the luminosity
of the iron line should decrease and disappear during the
re-burst, consistent with the observations of iron line(Piro et
al. 1999).

It is noticed that we assume the delayed energy is almost
isotropic. The energy obtained by the disk is
$E_{disk}=\frac{\Omega_{d}}{4\pi}E_{del}$, $\Omega_{d}$ is the
solid angle subtended by the fallback disk as observed at the
location of central engine. For the advection disk, H/R $\approx$
0.77(Narayan et al. 2001). We can get $E_{d}\approx 0.37 E_{del}$.
For GRB970508, the open half angle of the GRB collimated jet is
$16.7^{\circ}$(Frail et al. 2001). So the energy obtained by the
disk is $E_{d}\sim 60E_{0}$. Even if only $10\%$ fraction of the
energy was reflected by the disk(e.g. Zycki et al. 1994), It is
about 3$E_{0}$ reflected by one surface of the disk, that is
sufficient for the line emission production.

\section{Discussion and Conclusions }

It is found that the energy contained in the illuminating
continuum which is responsible for the line production is much
higher than that of the collimated main GRBs. Here we model that
the energy obtained from the delayed injection energy, higher than
that of the collimated $\gamma$-ray bursts, illuminates the
fallback disk which is formed after the supernovae explosion,
photoionizes the disk region of $\tau=1$, then produces the
observed iron line feature.

\textbf{In our model the delayed energy comes from the central
engine, which can be the magnetic energy from the declining
magnetic field of the superpulsar(Rees \&
M$\acute{e}$sz$\acute{a}$ros 2000) or the magnetic dipole
radiation of the magnetar(Dai \& Lu 1998). It could be primarily
in a magnetically driven relativistic wind(which would be
super-Eddington). The magnetized wind would develop a shock before
encountering the disk. The nonthermal electrons would be
accelerated behind the shock in the outflow material. The
shock-accelerated electrons could cool promptly, and would yield a
power-law X-ray continuum. It is similar to what has been proposed
by Rees and M$\acute{e}$sz$\acute{a}$ros(2000). This X-ray
continuum illuminates the fallback disk and produces the iron
line. The surface of the disk can be accelerated outward by this
super-Eddington flux of the illuminating continuum(e.g. Vietri et
al. 2001), so usually a outward velocity can be seen in the
lines(e.g. Reeves et al. 2002). In our model, the delayed energy
emission and the GRBs could come from different physical
processes. The delayed energy could be from magnetic wind of the
magnetar, so the energy emission is almost isotropic. However,
from the observation of the GRBs, the GRB prompt emission should
be intrinsically collimated. So there should exist a transition
from a collimated to an uncollimated energy release in the
engine.}

The Injection Energy must be higher than that of the initial main
burst so that the effect can be observed in the GRB
afterglows(Cohen \& Piran 1999;Zhang \&
M$\acute{e}$sz$\acute{a}$ros 2001). For GRB970508, at
$t\sim3\times 10^{3}$s after the GRB, the delayed illuminating
continuum decays to be $10^{47}erg s^{-1}$, and the ionized iron
ion recombines, Fe line appears and the line luminosity is about
$10^{44}-10^{45}erg s^{-1}$. at the time $t\sim 6\times 10^{4}$s,
the re-burst emerges. And after that time, the luminosity of the
delayed illuminating continuum decays to be less than $10^{46}erg
s^{-1}$, the iron line would decrease and disappear. The duration
of the Fe line $t_{d}\sim 10^{4}-10^{5} s$, longer than the
cooling time of thermal disk $t_{cool}\sim 10^{-4}n_{19}^{-1}$s.
All above are consistent with the observations of the Fe line in
GRB970508 X-ray afterglow.

The re-burst phenomenon has not been observed in GRB991216
afterglow. It can be explained that the Injection Energy is as
much as or less than that of the main burst, or the re-burst was
missed in the observation though it had happened. In the former
case, the line duration should be  $10^{4}$ s or less, consistent
with what has been observed, adopting the energy of the main burst
E$\sim10^{51}$ergs(Bloom et al. 2003).

\textbf{Above scenario is based on the supranova model in which
the time delay between the SN explosion and the $\gamma$-ray
bursts can be several months or even longer. Our model supports
the  'supranova model' because it must have enough time to let
nickel decay to iron (about $10^{-3}M_{\odot}$ Fe in the disk). In
our model we assume the time between the SN explosion and the GRB
is about 100 days, so the fallback disk that we consider has
evolved for about 100 days after the SN explosion. In this case,
the lines and the SN bump can not be seen in the same events. }

\textbf{In our model, the disk was in place before the GRB
occurred, so it may produce a high level of pre-GRB activity of
the source.}

\textbf{Different values of the ionization parameter, $\xi$, could
produce the different reflection spectra. When $\xi\sim 10^{2}$,
the spectra will show luminous lines from light metals and a
depressed $K_{\alpha}$ iron line; while $10^{3}<\xi<10^{5}$, a
luminous iron line will be observed in the spectra(Lazzati et al.
2002). In our model the fallback disk with different properties,
such as different of number density of the electrons in the disk
surface and the outermost radius, will have different ionization
parameter. In this paper, $\xi\sim 2.5\times10^{3}$, so luminous
$K_{\alpha}$ iron line can be observed in the spectra. }

Vietri et al.(1999) has suggested a thermal model that a
relativistic fireball associated with the GRB might hit the
pre-GRB supernova remnant within $\sim$ $10^{3}$s and heat the
ejecta to T$\sim$ $10^{7}$- $10^{8}$ K. At such temperature the
plasma emission shows thermal bremsstrahlung emission as well as
iron line emission. In their model the thermal bremsstrahlung and
recombination continuum from the thermal disk can account for the
re-burst observed in GRB970508 and GRB970828. While in our model,
the delayed injection energy from the central engine after the
main burst, more than the energy of the main burst accounts for
the re-burst and produces the iron line emission.

Our model is also different from the decaying magnetar model in
which Rees \& M$\acute{e}$sz$\acute{a}$ros(2000) suggested that
iron line could be attributed to the interaction of a continuing
but decaying postburst relativistic outflow from the central
engine with the progenitor stellar envelope at distances less than
a light-hour. In their model bumps should be found in less than
several hours after the GRBs(Gao \& Wei 2004).

In conclusion, We suggest that the delayed injection energy that
causes the re-burst in the GRB afterglow, illuminates the fallback
disk which is formed after the supernovae explosion, photonionizes
the fallback disk, then produces iron line feature. This scenario
can well explain the production of the re-burst and the emission
lines, that can be tested by the observations of SWIFT satellite
in the near future.

 \acknowledgments
We are very grateful to the referee for several important comments
that improved this paper. This work is supported by the National
Nature Science Foundation(grants 10233010,and 10225314) and the
National 973 Project on Fundamental Researches of China(NKBRSF
G19990745).

\newpage

\begin{figure}
\includegraphics[angle=0,scale=.50]{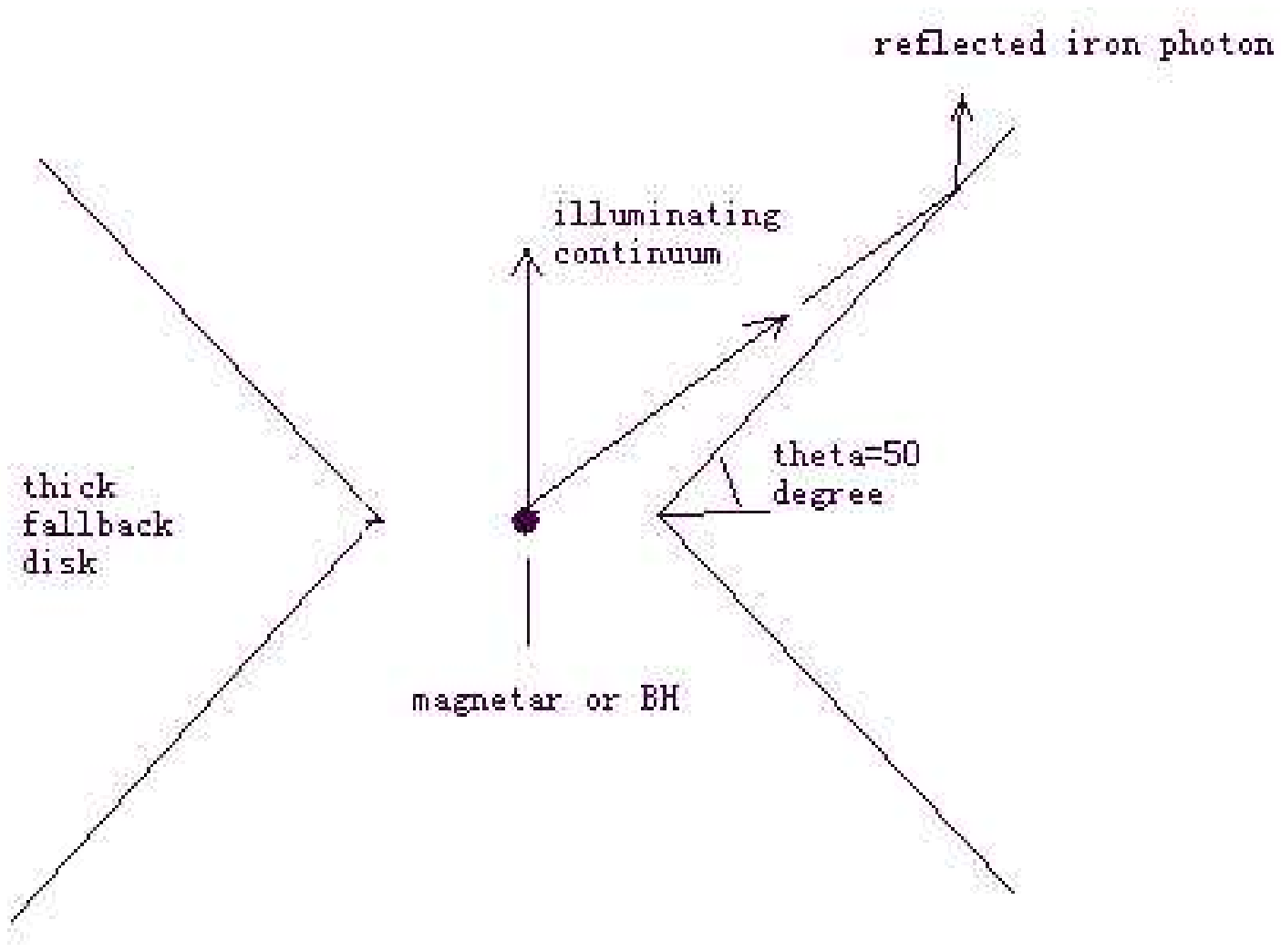}
\caption{This is just a cartoon picture of the geometry of the
fallback disk. The central engine is a magnetar or a black hole. }
\end{figure}

\newpage

\end{document}